\documentclass[12pt,aps,prd,superscriptaddress,showpacs,longbibliography,floatfix,nofootinbib]{revtex4-1}

\usepackage[utf8]{inputenc}
\usepackage{slashed}

\usepackage{color}
\usepackage{graphicx}   
\usepackage{bm}
\usepackage{amsmath}
\usepackage{amsfonts}
\usepackage{eufrak}
\usepackage{hyperref}

\newcommand{\be}{\begin{equation}}
\newcommand{\ee}{\end{equation}}
\newcommand{\ba}{\begin{eqnarray}}
\newcommand{\ea}{\end{eqnarray}}

\begin{document}

\title{On the negative coupling O(N) model in 2d at high temperature}

\author{Paul Romatschke}
\affiliation{Department of Physics, University of Colorado, Boulder, Colorado 80309, USA}
\affiliation{Center for Theory of Quantum Matter, University of Colorado, Boulder, Colorado 80309, USA}

\begin{abstract}
  In this work, I consider N-component scalar quantum field theory in two dimensions interacting with an upside-down quartic potential. Working in the large N limit, the model can be solved non-perturbatively using the saddle-point method for sufficiently strong negative coupling. At high temperature, the O(N) model dimensionally reduces to ${\cal PT}$-symmetric quantum mechanics, for which powerful non-perturbative solution methods exist. It is found that the solution from quantum mechanics can be matched by the saddle-point method in quantum field theory when allowing for saddles beyond the principal Riemann sheet. I show that saddle points on non-principal Riemann sheets lead to a fully consistent solution of the 2d negative-coupling O(N) model for all temperatures. 
\end{abstract}

\maketitle

\section{Introduction}

Most textbooks on continuum quantum field theory pretend that nature can be approximated as classical physics plus small perturbations in the coupling. While this approach may work well for many applications, it ignores the possibility that there are natural phenomena that are inherently quantum, meaning that they do not possess a viable expansion around a classical solution\footnote{A well-known example is the hydrogen atom, which does not have a classically stable ground state.}. 

Negative coupling quantum field theory (meaning quantum field theory with an upside-down potential) is a particular example for a theory that does not exist classically, but nevertheless may be quantum mechanically stable\footnote{Negative coupling quantum mechanics is a special case of ${\cal PT}$-symmetric quantum mechanics, which is known to exhibit positive definite Hamiltonian spectra and unitary time evolution \cite{Bender:1998ke,Bender:2019cwm}. Many predictions from ${\cal PT}$-symmetric quantum mechanics have been experimentally verified, cf. Refs.~\cite{Musslimani:2008zz,Nasari:2022dgn}.}. Historically dismissed as ``nonsense'' because of the absence of a classical ground state \cite{Coleman:1973sx}, in four dimensions negative coupling scalar quantum field theory has several attractive features such as asymptotic freedom \cite{Symanzik:1973hx}, unlike its positive coupling counterpart which is known to be quantum-trivial \cite{Aizenman:2019yuo}.

Studies of negative coupling quantum field theory are more difficult than ordinary quantum field theory because most of the tools for quantum field theory were built for theories which possess a classically stable ground state. For negative-coupling quantum mechanics numerical methods such as Hamiltonian diagonalization are available \cite{Bender:1998ke,Lawrence:2023woz}, but these approaches have not been generalized to quantum field theory (see, however, Ref.~\cite{diego}). Numerical approaches based on lattice discretization of the Euclidean path integral are possible \cite{Romatschke:2023fax}, but the absence of a classical ground state renders importance sampling techniques ineffective, thereby relegating available simulations to small lattice volumes. Nevertheless,  alternative approaches for negative coupling quantum field theory on the lattice could prove successful in the future \cite{Aarts:2013fpa,Lawrence:2022afv,Giachello:2024wqt,Boguslavski:2024yto}. Besides lattice discretization, the numerical bootstrap approach \cite{Barrat:2024fwq,Lawrence:2024mnj} also seems a promising direction for studying negative coupling quantum field theory.

For the present work, I will focus on yet another non-perturbative approach to studying negative coupling quantum field theory, namely semi-analytic expansion in a large number of scalar field components. This so-called ``large N'' expansion is a well-established technique that is known to provide systematically improvable quantitative results for positive-coupling quantum field theory with scalars and fermions \cite{ma1973critical,Klebanov:2002ja,Moshe:2003xn,veillette2007large,Romatschke:2019ybu,Pinto:2020nip,Grable:2022swa,Romatschke:2023ztk}. In four dimensions these quantum field theories generically predict a negative coupling in the ultraviolet \cite{Abbott:1975bn,Linde:1976qh,Romatschke:2022jqg,Grable:2023paf,Weller:2023jhc,Lyu:2024elz}(see, however, Refs.~\cite{Berges:2023rqa,Berges:2024ydj}), which nevertheless appears to be harmless as far as the thermodynamically preferred quantum vacuum is concerned \cite{Parisi:1975im,Romatschke:2023sce}. The resulting asymptotically free scalar quantum field theory in four dimensions holds the promise of generating electroweak masses while not suffering from many of the known problems of the Higgs mechanism \cite{Romatschke:2024hpb}.

However, as already pointed out in Ref.~\cite{Romatschke:2022jqg}, while the large N vacuum of four-dimensional large N models appears to be non-perturbatively stable, at sufficiently high temperature the large N saddle point solutions become complex, creating considerable difficulties for physics interpretation. While a conjecture put forward in Ref.~\cite{Ai:2022csx} suggested that the real part of the free energy should be taken in such cases, this conjecture subsequently was shown to not be true in general \cite{Lawrence:2023woz,Kamata:2023opn,Kamata:2024tyb}.

The present work is motivated by the question of what happens to negative-coupling quantum field theory in the high temperature limit. For simplicity, in this work I choose to study large N scalar quantum field theory in 1+1d Euclidean dimensions. This has several advantages: first, as will be shown below, the two dimensional negative coupling field theories exhibit the same structure that was found in four dimensions: stable vacuum solutions at low temperature and complex saddles at high temperature; second, the low dimensionality of these quantum field theories probably allow independent numerical studies, e.g. based on lattice discretization \cite{Romatschke:2023fax,Giachello:2024wqt}. And third, in the high temperature limit the length of the thermal circle shrinks to zero, so that the theory can be mapped onto ordinary quantum mechanics for which powerful solution techniques exist.

In the following sections, I will describe the setup of the problem, the high-temperature limit of the quantum field theory and the dimensional reduction to quantum mechanics. This will be followed by the solution of the quantum mechanical theory, from which a resolution of the complex saddle point problem will be identified. In the last section I will apply this resolution directly to the two-dimensional quantum field theory at any temperature, and show that thermodynamic stability at leading order in the large N limit also implies dynamic stability at next-to-leading order in the large N limit before concluding in the final section.

\section{Setup of the Problem}

Let me consider the path integral for the Euclidean partition function for the O(N) model in 1+1 dimensions:
\be
Z=\int {\cal D}\vec{\phi} e^{-S_E}\,,\quad S_E=\int_0^\beta d\tau \int dx\left[\frac{1}{2}\partial_\mu \vec{\phi}\partial_\mu \vec{\phi}+\frac{m_B^2\vec{\phi}^2}{2}+\frac{\lambda}{N}\left(\vec{\phi}^2\right)^2\right]\,,
\label{start}
\ee
where $\vec{\phi}=\left(\phi_1,\phi_2,\ldots,\phi_N\right)$ and $m_B$ is the bare mass and $\beta=\frac{1}{T}$ is the inverse temperature of the system. Using the mathematical identity
\be
\label{identity}
e^{-\frac{\lambda}{N}\left(\vec{\phi}^2\right)^2}=\int dz e^{-iz \vec{\phi}^2-\frac{N z^2}{4\lambda}}\,,
\ee
where $z(x)$ is an auxiliary field to rewrite $S_E$, the fields $\vec{\phi}$ enter the action quadratically and can formally be integrated out. The effective action thus becomes
\be
\label{action}
S_{\rm eff}=\frac{N}{2}{\rm Tr}\ln\left[-\Box+m_B^2+i z(x)\right]-N \int d^2x  \frac{(iz(x))^2}{16\lambda}\,.
\ee
In the following, I am interested in the theory at \textit{negative coupling}
\be
\label{rep}
\lambda=-g\,,\quad g>0\,.
\ee
As has been shown in Ref.~\cite{Weller:2023jhc}, auxiliary field action for the negative-coupling theory is given by (\ref{action}) with the simple replacement (\ref{rep}).

In the large N limit, the partition function $Z$ can be evaluated from the saddle points of the action. Using the well-known (and for positive coupling, well-tested) machinery of saddle-point expansions, one writes
\be
z(x)=z_0+\xi(x)\,,
\ee
with constant $z_0$ and where the contributions from the fluctuations $\xi(x)$ are known to be $1/N$ suppressed in the large N limit \cite{Romatschke:2023ztk}. To leading order in large N, one has
\be
\ln Z=-S_{\rm eff}[z_0=\bar z]+{\cal O}\left(\frac{1}{N}\right)\,,
\ee
where the saddle point(s) $\bar z$ are the solutions of the saddle-point condition
\be
\frac{1}{N}\frac{d S_{\rm eff}[z_0]}{dz_0}=0=\frac{iz_0}{8g}+\frac{T}{2}\sum_n \int \frac{dk}{2\pi}\frac{1}{\omega_n^2+m_B^2+k^2+iz_0}\,,
\label{saddle}
\ee
where $\omega_n=2 \pi n T$ are the bosonic Matsubara frequencies and results from field theory at finite temperature $T$ have been used to rewrite the trace over the logarithm of an operator, cf. Refs.~\cite{Laine:2016hma,Romatschke:2023ztk}. The sum in (\ref{saddle}) can be done in closed form 
\be
\label{sum1}
\frac{T}{2}\sum_n \int \frac{dk}{2\pi}\frac{1}{\omega_n^2+m_B^2+k^2+iz_0}=\int \frac{dk}{2\pi}\frac{1+2 n_B\left(\sqrt{k^2+m_B^2+iz_0}\right)}{4 \sqrt{k^2+m_B^2+iz_0}}\,,
\ee
where $n_B(x)=\frac{1}{e^{\beta x}-1}$ is the Bose-Einstein distribution.

\subsection{Zero temperature limit}
In the zero-temperature limit, $n_B(x)\rightarrow 0$ and integral in (\ref{sum1}) can be evaluated using dimensional regularization for $d=2-2\varepsilon$. Using this result in the saddle-point condition (\ref{saddle}), one has  
\be
T=0:\quad \frac{iz_0}{8g}+\frac{1}{8\pi}\left[\frac{1}{\varepsilon}+\ln \frac{\bar\mu^2}{m_B^2+iz_0} \right]=0\,,
\ee
where $\bar\mu$ is the $\overline{\rm MS}$ scheme parameter. Shifting $iz_0\rightarrow iz_0-m_B^2$ one can renormalize the saddle-point condition by choosing
\be
\label{renorm}
\frac{m_B^2}{8g}-\frac{1}{8\pi \varepsilon}=\frac{m_R^2(\bar\mu)}{8g}\,,
\ee
with the running mass parameter
\be
\frac{m_R^2(\bar\mu)}{8g}=\frac{1}{\pi}\ln \frac{\bar\mu^2}{\Lambda_{\overline{\rm MS}}^2}\,,
\ee
with $\Lambda_{\overline{\rm MS}}$ the $\overline{\rm MS}$  scale parameter. The renormalized saddle-point condition at zero temperature is
\be
T=0:\quad \frac{iz_0}{g}+\frac{1}{\pi}\ln \frac{\Lambda_{\overline{\rm MS}}^2}{iz_0}=0\,,
\ee
which has two solutions
\be
\label{zeroTsolutoins}
T=0:\quad i\bar z_+=-\frac{g}{\pi}{\rm W}_0\left(-\frac{\Lambda_{\overline{\rm MS}}^2 \pi}{g}\right)\,,\quad i\bar z_{-}=-\frac{g}{\pi}{\rm W}_{-1}\left(-\frac{\Lambda_{\overline{\rm MS}}^2 \pi}{g}\right)\,,
\ee
where $W_k(x)$ denotes the Lambert $W$ function of branch $k$. The nature of the solution changes depending on the value of the coupling constant. Defining
\be
g_{\rm crit}\equiv \Lambda_{\overline{\rm MS}}^2 \pi e^1,
\ee
the zero-temperature saddles (\ref{zeroTsolutoins}) are real and positive and obey $i\bar z_{-}>i\bar z_+$ for $g\geq g_{\rm crit}$, but are complex conjugate pairs $i\bar z_{-}=\left(i\bar z_+\right)^*$ for $g<g_{\rm crit}$. In the limiting case
\be
g=g_{\rm crit}:\quad i\bar z_{-}=i\bar z_+=e \Lambda_{\overline{\rm MS}}^2\,,
\ee
the solutions become degenerate.

For $g\geq g_{\rm crit}$, the action for both solutions (\ref{zeroTsolutoins}) is real, and the free energy density is given by
\be
\Omega\equiv \frac{S_{\rm eff}[\bar z]}{\beta L}\,,
\ee
with $L$ the spatial volume. $\Omega$ depends on the value of the saddle-point $\bar z$ and therefore the different saddle points can be interpreted as characterizing different phases of the theory, with the preferred phase given by the saddle that has the lower free energy. Evaluating the renormalized action $S_{\rm eff}[\bar z]$, one finds
\be
T=0:\quad \frac{\Omega[z_0]}{N}=\frac{(iz_0)^2}{16g}+\frac{iz_0}{8\pi}\ln \frac{e^{1}\Lambda_{\overline{\rm MS}}^2}{iz_0}\,,
\ee
and the lower free energy solution is given by the saddle $\bar z_{-}$:
\be
T=0: \quad \frac{\Omega[\bar z_{-}]}{N}=-\frac{(i\bar z_{-})^2}{16g}+\frac{i\bar z_{-}}{8\pi}\,.
\ee
Note that in the critical coupling case
\be
g=g_{\rm crit}:\quad \Omega[\bar z_{-}]=\frac{N e\Lambda_{\overline{\rm MS}}^2}{16 \pi}\,.
\ee

\subsection{Complex saddles at finite temperature}

At finite temperature, one evaluates (\ref{sum1}) by using
\be
\int_0^\infty \frac{dk}{2\pi}\frac{n_B(\sqrt{k^2+iz_0})}{\sqrt{k^2+iz_0}}=\sum_{n=1}^\infty\int_0^\infty \frac{dk}{2\pi} \frac{e^{n\beta \sqrt{k^2+iz_0}}}{\sqrt{k^2+iz_0}}=\sum_{n=1}^\infty \frac{K_0\left(n \beta \sqrt{iz_0}\right)}{2\pi}\,,
\ee
where $K_0(x)$ denotes a modified Bessel function of the second kind. The saddle-point condition (\ref{saddle}) at finite temperature therefore becomes
\be
\label{saddle1}
0=\frac{iz_0}{8g}+\frac{1}{8\pi}\ln\frac{\Lambda_{\overline{\rm MS}}^2}{iz_0}+\sum_{n=1}^\infty \frac{K_0\left(n \beta \sqrt{iz_0}\right)}{2\pi}\,,
\ee
and solutions can be found numerically.

For $g>g_{\rm crit}$ and  small values of the temperature $T\ll \Lambda_{\overline{\rm MS}}$, one finds two solutions $\bar z_{\pm}$ which are analytically connected to the zero-temperature solutions (\ref{zeroTsolutoins}). However, above some temperature $T>T_{c1}$, the finite temperature solutions $i\bar z_{\pm}$ become complex-conjugate pairs, in complete analogy to what was found for the case of the O(N) model in four dimensions \cite{Weller:2023jhc,Romatschke:2022jqg,Romatschke:2023sce}. The complex saddle-points for $T>T_{c1}$ also mean that the free-energy $\Omega[i\bar z]$ for those saddles is complex, which challenges a physical interpretation of these saddle-points as physical vacua. For the four-dimensional O(N) model, Ref.~\cite{Romatschke:2022jqg} made an appeal to the ABS conjecture \cite{Ai:2022csx}, which posits that the physical free energy is the real part of the complex-valued action. However, there are at least two problems with this interpretation: first, there is mounting evidence in quantum mechanics that the ABS conjecture does not hold in general \cite{Lawrence:2023woz,Kamata:2023opn,Kamata:2024tyb,Chen:2024ynx}. Second, the saddle-point solutions $i\bar z$ appear as the mass parameter in the propagator of the fields $\vec{\phi}$, cf. Eq.~(\ref{action}). Complex masses imply that field propagators have poles in the unphysical frequency half-plane, which usually indicates the presence of an instability in the system (see Refs.~\cite{Romatschke:2003ms,Arnold:2003rq,Romatschke:2005pm,Mrowczynski:2016etf} for an explicit example of such an instability for non-Abelian plasmas).  Therefore, even if the ABS conjecture were to hold, the seemingly unavoidable presence of these instabilities further challenges the physical interpretation of these complex saddles.

\section{High temperature limit}

In order to study the phase structure of the theory when complex saddle points are present, it is advisable to focus on the parameter region that is simplest to control. For this reason, I will now investigate what happens in the high temperature limit of the theory, for which the length of the thermal circle shrinks to zero $\beta\rightarrow 0$. The high-temperature limit therefore exhibits dimensional reduction, which is a well-developed and much used approach in conventional thermal field theory \cite[Chap. 6.2]{Laine:2016hma}.

For high temperature, it is possible to extract the analytic behavior of the thermal sum (\ref{sum1}) by using the high-temperature expansion from \cite[(2.90)]{Laine:2016hma}. This is done by isolating the Fourier zero mode $n=0$ in (\ref{sum1}) and expanding the remaining integrand for $n\neq 0$ in powers of $\omega_n^2\gg k^2$. One thus finds the alternative representation for the saddle-point condition (\ref{saddle1})
\be
\label{saddle2}
0=\frac{iz_0}{8g}+\frac{ T}{4\sqrt{iz_0}}+\frac{1}{8\pi}\ln\frac{\Lambda_{\overline{\rm MS}}^2 e^{2\gamma_E}}{(4 \pi T)^2}+T \sum_{l=1}^\infty \frac{(-1)^l (iz_0)^{2l}}{\sqrt{4\pi}}\frac{\Gamma\left(l+\frac{1}{2}\right)}{\Gamma\left(l+1\right)}\frac{\zeta(2l+1)}{(2\pi T)^{2l+1}}\,,
\ee
where $\zeta(s)$ denotes the Riemann zeta function.
Comparing (\ref{saddle1}), (\ref{saddle2}), it should be remarked that  the logarithm of $iz_0$ in (\ref{saddle1}) has canceled against the sum over Bessel functions to become a logarithm of temperature in (\ref{saddle2}). At high temperature, the saddle-point condition is well approximated by the first two terms in (\ref{saddle2}), and one readily identifies the solutions
\be
\label{zhighT}
T\rightarrow \infty:\quad i\bar z_\pm=\left(-2 T g\right)^{\frac{2}{3}}=e^{\pm \frac{2 i \pi}{3}}(2 T g)^{\frac{2}{3}}\,.
\ee
In analogy to the numerically obtained solutions to (\ref{saddle1}), (\ref{zhighT}) are complex-valued and correspond to a pair of complex-conjugate solutions.

From (\ref{action}), the renormalized free-energy density at finite temperature is given by
\ba
\frac{\Omega}{N}=\frac{T}{2} \sum_n \int \frac{dk}{2\pi}\ln\left(\omega_n^2+k^2+iz_0\right)+\frac{(iz_0)^2+2 iz_0 m_B^2+m_B^4}{16 g}\,,
\ea
which after renormalization (\ref{renorm}) and subtracting a divergent term $\frac{m_B^4}{16 g}$ becomes
\be
\frac{\Omega}{N}=\frac{iz_0}{8\pi}\ln\frac{\Lambda_{\overline{\rm MS}}^2 e^1}{iz_0}-\sum_{n=1}^\infty\frac{T \sqrt{iz_0}K_1\left(n \beta \sqrt{iz_0}\right)}{\pi n} +\frac{(iz_0)^2}{16 g}\,.
\ee
Using the asymptotic form of the modified Bessel function in the high temperature limit $\beta\rightarrow 0$, one finds
\be
T\rightarrow \infty: \quad \frac{\Omega}{N}=-P_{SB}(T)=-\frac{\pi T^2}{6}+\ldots\,,
\ee
where $P_{SB}(T)=\frac{\pi T^2}{6}$ is the Stefan-Boltzmann pressure in two dimensions. An alternative expression for the renormalized free-energy density is to integrate (\ref{saddle2}) with respect to $iz_0$, with the integration constant determined such that the free energy matches the Stefan-Boltzmann value in the high temperature limit. This procedure gives
\be
\label{Shigh}
\frac{\Omega}{N}=-\frac{\pi T^2}{6}+\frac{T \sqrt{iz_0}}{2}+\frac{(iz_0)^2}{16 g}+\frac{iz_0}{8\pi}\ln \frac{\Lambda_{\overline{\rm MS}}^2}{(4 \pi T)^2}+T\sum_{l=1}^\infty \frac{(-1)^l (iz_0)^{2l+2}}{\sqrt{4\pi}}\frac{\Gamma\left(l+\frac{1}{2}\right)}{\Gamma\left(l+2\right)}\frac{\zeta(2l+1)}{(2\pi T)^{2l+1}}\,.
\ee
Naively inserting the complex saddle-point solutions (\ref{zhighT}) into (\ref{Shigh} leads to
\be
\label{naive}
T\rightarrow \infty:\quad \frac{\Omega}{N}=-\frac{\pi T^2}{6}+\frac{3 T^{\frac{4}{3}}(2g)^{\frac{1}{3}}}{8}e^{\pm \frac{i\pi}{3}}+\ldots,,
\ee
which is complex-valued.

The propagator for the fields $\vec{\phi}$ is given by 
\be
\langle\phi_i(x)\phi_j(0)\rangle\equiv\delta_{ij}G(x)=T\sum_n \int \frac{dk}{2\pi}
\frac{\delta_{ij} e^{i\omega_n \tau+i k x}}{\omega_n^2+k^2+iz_0}\,.
\ee
In the high temperature limit, one has in momentum space
\be
\label{Ght}
G(n=0,k)=\frac{1}{k^2+(-2 g T)^\frac{2}{3}}\,,
\ee
where (\ref{zhighT}) again has been used.

\subsection{Effective dimensionally reduced theory}

In the high temperature limit, the length of the thermal circle tends to zero, and the one-dimensional effective action after rescaling $\vec{\phi}\rightarrow T \vec{\phi}$ in this limit becomes
\be
S_{1d}=\int dx \left[\frac{1}{2}\partial_x\vec{\phi}\partial_x \vec{\phi}+\frac{m_{\rm 1d}^2 \vec{\phi}^2}{2}-\frac{g T}{N}\left(\vec{\phi}^2\right)^2\right]\,,
\ee
which is a quantum-mechanical theory. Here $m_{\rm 1d}^2$ is an effective parameter that can be fixed by calculating correlation functions in both the dimensionally-reduced and original theory. Using (\ref{identity}) from above gives
\be
\label{S1d}
S_{\rm 1d}=\int dx \left[\frac{\vec{\phi}}{2}\left[-\partial_x^2+m_{\rm 1d}^2+iz_{\rm 1d}\right] \vec{\phi}+\frac{N (iz_{\rm 1d})^2}{16 g}\right]\,,
\ee
where $z_{\rm 1d}$ is a solution to the 1d saddle-point condition
\be
\label{effgap}
iz_{\rm 1d}=-\frac{2 g T}{\sqrt{iz_{\rm 1d}+m_{\rm 1d}^2}}\,.
\ee
Calculating the $\vec{\phi}$ propagator in the 1d effective theory then gives
\be
\langle \phi_i(x)\phi_j(0)\rangle=\delta_{ij}G_{1d}(x)=\int \frac{dk}{2\pi}\frac{e^{i k x}}{k^2+m_{\rm 1d}^2+i z_{\rm 1d}}\,,
\ee
which must match (\ref{Ght}). As a consequence, one finds that matching the propagator as well as using the gap equation (\ref{effgap}) fixes the effective mass parameter in the dimensionally reduced theory as
\be
m_{\rm 1d}^2=0\,.
\ee

The partition function for the effective theory is then given by
\be
\ln Z_{\rm 1d}=-L E_0\,,
\ee
where $E_0$ is the ground-state energy of the quantum mechanical action $S_{\rm 1d}$ and $L$ is the spatial volume. The free-energy density in the \textit{original} field theory can be related to the dimensionally reduced theory as
\be
\Omega_{2d}=-\frac{\ln Z_{\rm 1d}}{\beta L}=T E_0\,.
\ee
It is easy to see that this procedure only captures the deviations from the 2d free theory Stefan-Boltzmann limit, so that the exact relation between the free-energy density in the original theory and the dimensionally reduced theory is given by
\be
\label{o2d}
T\rightarrow \infty: \quad \Omega=-\frac{N T^2 \pi}{6}+T E_0+\ldots\,.
\ee

Naively, one could try to obtain the ground state energy $E_0$ by noting that (\ref{S1d}) is just quantum mechanics for an anharmonic quartic oscillator with a  funny coupling constant. (Falsely) assuming that one can take the positive coupling large N result for quantum mechanics (see e.g. \cite[Eq. (25)]{Romatschke:2023ztk}), and flip the sign of the coupling one would get
\be
\label{funny}
E_0=\frac{3 N}{8} (-2 g T)^{\frac{1}{3}}+{\cal O}(N^0)\,, \quad \Omega=-\frac{T^2 \pi}{6}+\frac{3 N T^{\frac{4}{3}}}{8} (-2 g)^{\frac{1}{3}}+{\cal O}(N^0)\,,
\ee
matching (\ref{naive}). Using the principal branches of the root function, the ground-state energy $E_0$ for the negative coupling theory defined by (\ref{S1d}) would appear to be complex-valued.

The error made in this procedure is that the ground-state energy for the action (\ref{S1d}) is \textit{not} given by flipping the sign of the coupling in the result obtained for the positive-coupling theory calculation \cite{Lawrence:2023woz,Kamata:2023opn}. The correct ground state energy for the quantum mechanical system defined by the action (\ref{S1d}) has been obtained in Ref.~\cite{Bender:1998ke} for the case N=1, showing that $E_0$ is real, unlike the naive result (\ref{funny}). 

Therefore, to correctly calculate $E_0$ for (\ref{S1d})  one needs to properly solve the quantum mechanics problem.

  \subsection{Hermitian Equivalent of Large N Upside Down Quartic Oscillator}

  One starts with the discretized path-integral for the N-component upside-down quartic oscillator:
  \be
  Z_{1d}=\int \prod_x \frac{d\phi_1(x)d\phi_2(x)\ldots d\phi_N(x)}{(2 \pi \varepsilon)^{N/2}} \exp\left[-\sum_{i=1}^N \frac{(\phi_i(x+\varepsilon)-\phi_i(x))^2}{2\varepsilon}+\frac{\varepsilon g}{N}\left(\sum_{i=1}^N \phi_i^2(x)\right)^2\right]\,.
  \ee
  Using the parametrization
  \be
  \phi_i(x)=-2 i \sqrt{1+i \psi_i(x)}
  \ee
  from Ref.~\cite{Jones:2006qs}, one can use the technique outlined in Ref.~\cite{Bender:2006wt} to rewrite
  \be
  \frac{(\phi_i(x+\varepsilon)-\phi_i(x))^2}{2}=\frac{\left(\psi_i(x+\varepsilon)-\psi_i(x)\right)^2}{2 f_i^2(x)}\,,\quad f_i(x)\equiv \frac{\sqrt{1+i \psi_i(x+\varepsilon)}+\sqrt{1+i \psi_i(x)}}{2}\,,
  \ee
  so that
  \be
  Z_{1d}=\int \prod_x \frac{d\psi_1(x)d\psi_2(x)\ldots d\psi_N(x)}{(2 \pi \varepsilon)^{N/2}} J(x) e^{-\sum_{i=1}^N \frac{(\psi_i(x+\varepsilon)-\psi_i(x))^2}{2\varepsilon f_i^2(x)}+\frac{16\varepsilon g}{N}\left(N+\sum_{i=1}^N i \psi_i(x)\right)^2}\,,
  \ee
  where for every site $x$ the Jacobian $J$ can be rewritten as \cite{Bender:2006wt}
  \ba
  J(x)&=&\prod_{i=1}^N \frac{1}{\sqrt{1+i \psi_i(x)}}=\prod_{i=1}^N\frac{1}{f_i(x)}\left(1+\frac{i (\psi_i(x+\varepsilon)-\psi_i(x))}{4 f_i(x) \sqrt{1+i \psi_i(x)}}\right)\,,\nonumber\\
  &\simeq&\prod_{i=1}^N\frac{1}{f_i(x)}\left(1+\frac{i (\psi_i(x+\varepsilon)-\psi_i(x))}{4 f_i^2(x)}+{\cal O}(\varepsilon^2)\right)\,,\nonumber\\
  &\simeq&\prod_{i=1}^N\int \frac{d p_i(x)}{\sqrt{2\pi/\varepsilon}}\left(1+\frac{\varepsilon p_i(x)}{4}\right)e^{-\frac{\varepsilon f_i^2(x)}{2} \left(p_i(x)-\frac{i (\psi_i(x+\varepsilon)-\psi_i(x))}{\varepsilon f_i^2(x)}\right)^2}\,.
  \ea

  Hence the partition function becomes
  \be
  Z_{1d}=\int \prod_x \left(\prod_{i=1}^N\frac{d\psi_i dp_i}{2 \pi}\right)  e^{-\sum_{i=1}^N\left[\frac{\varepsilon f_i^2(x) p_i^2(x)}{2} - i p_i(x)(\psi_i(x+\varepsilon)-\psi_i(x))-\frac{\varepsilon p_i(x)}{4}\right]+\frac{16\varepsilon g}{N}\left(N+\sum_{i=1}^N i \psi_i(x)\right)^2}\,.
  \ee
Replacing $f_i(x)=\sqrt{1+i \psi_i(x)}+{\cal O}(\varepsilon)$ and shifting all $N$ fields $\psi_i(x)\rightarrow \psi_i(x)-1$ leads to
   \be
  Z_{1d}=\int \prod_x \left(\prod_{i=1}^N\frac{d\psi_i dp_i}{2 \pi}\right)  e^{-\sum_{i=1}^N\left[\frac{\varepsilon i \psi_i(x) p_i^2(x)}{2} - i p_i(x)(\psi_i(x+\varepsilon)-\psi_i(x))-\frac{\varepsilon p_i(x)}{4}\right]-\frac{16\varepsilon g}{N}\left(\sum_{i=1}^N \psi_i(x)\right)^2}\,.
  \ee
  Using the product over sites $x$ to reshuffling the exponential gives
  \be
  Z_{1d}=\int \prod_x \left(\prod_{i=1}^N\frac{d\psi_i dp_i}{2 \pi}\right)  e^{-\sum_{i=1}^N\left[i\psi_i(x)\left(\frac{\varepsilon p_i^2(x)}{2} + p_i(x)-p_i(x-\varepsilon)\right)-\frac{\varepsilon p_i(x)}{4}\right]-\frac{16\varepsilon g}{N}\left(\sum_{i=1}^N \psi_i(x)\right)^2}\,.
  \ee
  Rewriting
  \be
  \psi_1(x)=\psi_1(x)-\sum_{i=2}^N \psi_i(x)
  \ee
  one can integrate over $\psi_1(x)$ which leads to
  \ba
  Z_{1d}&=&\int \prod_x \frac{dp_1}{8\sqrt{\pi g \varepsilon/N}}\left(\prod_{i=2}^N\frac{d\psi_i dp_i}{2 \pi}\right)\nonumber\\
  &&\times e^{-\frac{\varepsilon p_1(x)}{4}-\frac{N\varepsilon\left(\frac{p_1^4(x)}{4}+\dot p_1^2(x)\right)}{64 g}-\varepsilon \sum_{i=2}^N\left[i \psi_i(x)\left(\frac{p_i^2(x)-p_1^2(x)}{2} + \dot p_i(x)-\dot p_1(x)\right)-\frac{ p_i(x)}{4}\right]}\,,
  \ea
  where
  \be
  \dot p_i(x)\equiv \frac{p_i(x)-p_i(x-\varepsilon)}{\varepsilon}\,.
  \ee
  Noting that all remaining integrations over $\psi_i$, $i=2,\ldots,N$ lead to Dirac delta-functions, one has
  \ba
  Z_{1d}&=&\int \prod_x \frac{dp_1}{8\sqrt{\pi g \varepsilon/N}}\left(\prod_{i=2}^N d p_i \delta\left(\varepsilon \frac{p_i^2(x)-p_1^2(x)}{2} + \varepsilon \dot p_i(x)-\varepsilon \dot p_1(x)\right)\right)\nonumber\\
  &&\times e^{-\frac{N\varepsilon\left(\frac{p_1^4(x)}{4}+\dot p_1^2(x)\right)}{64 g}+ \sum_{i=1}^N\frac{\varepsilon p_i(x)}{4}}\,. 
  \ea
  Carefully implementing periodic boundary conditions, one finds that the delta functions fix all momenta $p_i(x)$, $i=2,\ldots,N$ as 
  \be
  p_i(x)=p_1(x)\,,
  \ee
  and there is only one overall non-trivial factor. One finds
\be
Z_{1d}=\int \left(\prod_x \frac{dp_1}{8\sqrt{\pi g \varepsilon/N}}\right) e^{-N \varepsilon\sum_x\left[\frac{\left(\frac{p_1^4(x)}{4}+\dot p_1^2(x)\right)}{64 g}- \frac{p_1(x)}{4}\right]-(N-1)\ln \left[\varepsilon \sum_x p_1(x)\right]}\,.
\ee
Rescaling
\be
p_1(x)=p(x) \sqrt{\frac{32 g}{N}}
\ee
leads to the result for the partition function in continuum notation
\be
Z_{1d}=\int {\cal D}p e^{-\int_0^\beta dx \left[\frac{  \dot p^2(x)}{2}+\frac{4 g}{N} p^4(x)  - \sqrt{2 g N} p(x)\right]-(N-1)\ln\left[\int_0^\beta dx p(x)\right]}\,. 
\ee
In the large N limit and zero temperature limit, the partition function localizes around the classical minimum of the potential
\be
\label{potty}
V(p)=\frac{4 g}{N}p^4-\sqrt{2 g N}p\,,
\ee
where the logarithmic term does not contribute at zero temperature. The minimum of the potential is found as
\be
p=\bar p=\frac{\sqrt{N}}{2^{\frac{7}{6}} g^{\frac{1}{6}}}\,,
\ee
so that
\be
\label{NLOE0}
Z=e^{-\beta V(\bar p)}=e^{-\beta E_0}\,,\quad E_0=-\frac{3 N (2g)^{\frac{1}{3}}}{8}\,.
\ee
Note that $E_0<0$ and that it corresponds to the \textit{non-principal root} in the solution (\ref{funny}).

To close this discussion on quantum mechanics, note that the ground state energy $E_0$ may be obtained numerically for any value of $N$ by diagonalizing the quantum mechanical Hamiltonian with potential (\ref{potty}). For reference, I list the results for $E_0$ for selected values of $N$ in table \ref{tab:one}.

\begin{table}
  \begin{tabular}{|c|c|c|c|c|c|c|c|}
    \hline
    N & 1 & 2 & 3 & 4 & 10 & $10^2$ & $10^3$\\
    \hline
    $\frac{E_0}{N g^{\frac{1}{3}}}$ & 0.9305 & 0.2231 & 0. & -0.11236 & -0.3227 & -0.4507 & -0.4709\\
    \hline
  \end{tabular}
  \caption{\label{tab:one} Values for the ground state energy $E_0/N$ in units of $g^{\frac{1}{3}}$ for the upside-down quartic oscillator with action (\ref{S1d}) for various values of N, obtained using numerical diagonalization. The value for $N=1$ matches that reported in Ref.~\cite{Bender:1998ke} using the normalization adopted in this work. Curiously, it seems that $E_0\simeq 0$ to within numerical precision for $N=3$.}
  \end{table}

\subsection{The proper high-temperature limit of the 2d O(N) model}

Using the result (\ref{NLOE0}) for the ground state energy $E_0$, one finds for the high-temperature limit and large N limit (\ref{o2d}) of the 2d O(N) model the result
\be
\label{dredO}
T\rightarrow \infty: \quad \Omega=-\frac{N T^2 \pi}{6}-\frac{3 N (2g)^{\frac{1}{3}} T^{\frac{4}{3}}}{8}+\ldots\,.
\ee
This result generalizes to finite N:
\be
T\rightarrow \infty: \quad \Omega=-\frac{N T^2 \pi}{6}+T E_0+{\cal O}\left(T^{\frac{2}{3}}\right)\,,
\ee
with $E_0$ given numerically for finite $N$ in Table \ref{tab:one}. In particular, I point out the curious result for $N=3$:
\be
T\rightarrow \infty: \quad \Omega=-\frac{N T^2 \pi}{6}+{\cal O}\left(T^{\frac{2}{3}}\right)\,,
\ee
whereas for $N=1$ the result is
\be
T\rightarrow \infty: \quad \Omega\simeq -\frac{T^2 \pi}{6}+0.9305 g^{\frac{1}{3}}T^{\frac{4}{3}}+{\cal O}\left(T^{\frac{2}{3}}\right)\,.
\ee
It will be interesting to test these semi-analytic results by direct numerical simulations of the negative-coupling 2d O(N) model through lattice simulations, potentially along the lines of Refs.~\cite{Romatschke:2023fax,Giachello:2024wqt}.

\section{Resolution of the Problem of Complex Masses}

Summarizing the last section, it is found that in the high-temperature limit, the large N physics of the two-dimensional O(N) model is correctly captured by allowing for saddle-point solutions away from the principal Riemann sheet.

This insight can be used to solve the large N limit of the O(N) model for any temperature, not only in the high temperature limit. Are there solutions to the saddle-point condition (\ref{saddle1}) away from the principal Riemann sheet?

In order to identify solutions, the saddle-point condition must be brought into a form that allows analytic continuation beyond the principal Riemann sheet. This can be accomplished by using the alternative form (\ref{saddle2}), expressing the Riemann zeta functions as $\zeta(s)=\sum_{n=1}^\infty n^{-s}$, and performing the sum over $l$ to find
\be
\label{saddle3}
0=\frac{iz_0}{8g}+\frac{ T}{4\sqrt{iz_0}}+\frac{1}{8\pi}\ln\frac{\Lambda_{\overline{\rm MS}}^2 e^{2\gamma_E}}{(4 \pi T)^2}+\frac{1}{4\pi}\sum_{n=1}^\infty\left(\frac{1}{\sqrt{n^2+\frac{iz_0}{(2 \pi T)^2}}}-\frac{1}{n}\right)\,.
\ee
By truncating the infinite sums in (\ref{saddle1}), (\ref{saddle3}), one can check numerically that both are identical expressions for $z_0$ on the principal Riemann sheet. To extend (\ref{saddle3}) beyond the principal Riemann sheet, I analytically continue
\be
\sqrt{iz_0}\rightarrow m\,,\quad m\in {\mathbb C}\,,
\ee
so that (\ref{saddle3}) takes the form
\be
\label{saddle4}
0=\frac{m^2}{8g}+\frac{ T}{4m}+\frac{1}{8\pi}\ln\frac{\Lambda_{\overline{\rm MS}}^2 e^{2\gamma_E}}{(4 \pi T)^2}+\frac{1}{4\pi}\sum_{n=1}^\infty\left(\frac{1}{\sqrt{n^2+\frac{m^2}{(2 \pi T)^2}}}-\frac{1}{n}\right)\,.
\ee
In the high temperature limit, only the first two terms of this expression are relevant, and one identifies the solution
\be
\label{Rzhigh}
T\rightarrow \infty: \quad m=-(2 T g)^{\frac{1}{3}}\,,
\ee
which indeed corresponds to the extension of (\ref{zhighT}) away from the principal Riemann sheet.

An extension for the free energy density that is valid beyond the principal Riemann sheet can likewise be obtained from (\ref{Shigh}) as
\be
\label{Shigh4}
\frac{\Omega}{N}=-\frac{\pi T^2}{6}+\frac{T m}{2}+\frac{m^4}{16 g}+\frac{m^2}{8\pi}\ln \frac{\Lambda_{\overline{\rm MS}}^2}{(4 \pi T)^2}+\frac{m^2}{2\pi}
\sum_{n=1}^\infty \left[\frac{(2 \pi T)^2}{m^2}\left(\sqrt{n^2+\frac{m^2}{(2 \pi T)^2}}-n\right)-\frac{1}{2n}\right]\,.
\ee
In the high-temperature limit, only the first three terms contribute, and using the saddle-point (\ref{Rzhigh}) one gets
\be
T\rightarrow \infty: \quad \frac{\Omega}{N}=-\frac{\pi T^2}{6}-\frac{3 T^\frac{4}{3} (2g)^{\frac{1}{3}}}{8}+\ldots\,,
\ee
matching (\ref{dredO}) from the dimensionally reduced theory.

However, (\ref{saddle4}) holds at any value of the temperature, so that one can search for saddle-point solutions directly in the two-dimensional quantum field theory. In particular, in the small-temperature limit, one recovers the solutions (\ref{zeroTsolutoins}) on the principal Riemann sheet:
\be
\label{lolli}
m_{-1}=\sqrt{i\bar z_{-}},\quad m_{0}=\sqrt{i\bar z_+}\,,
\ee
and in addition the solutions
\be
\label{lolli2}
m_{1}=-\sqrt{i\bar z_{-}},\quad m_{2}=-\sqrt{i\bar z_+}\,,
\ee
as well as
\be
m_{3}=-\frac{2 \pi T}{\ln \frac{\Lambda_{\overline{\rm MS}}^2 e^{2\gamma_E}}{(4 \pi T)^2}}\,.
\ee
Since $m_3$ vanishes in the zero temperature limit, one has
\be
T=0: \quad \frac{\Omega[m_3]}{N}=0\,.
\ee

\begin{figure}
  \includegraphics[width=.48\linewidth]{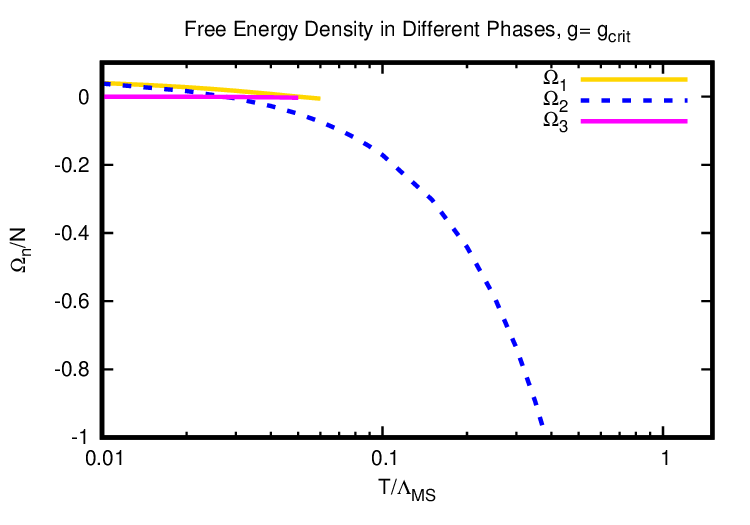}\hfill
  \includegraphics[width=.48\linewidth]{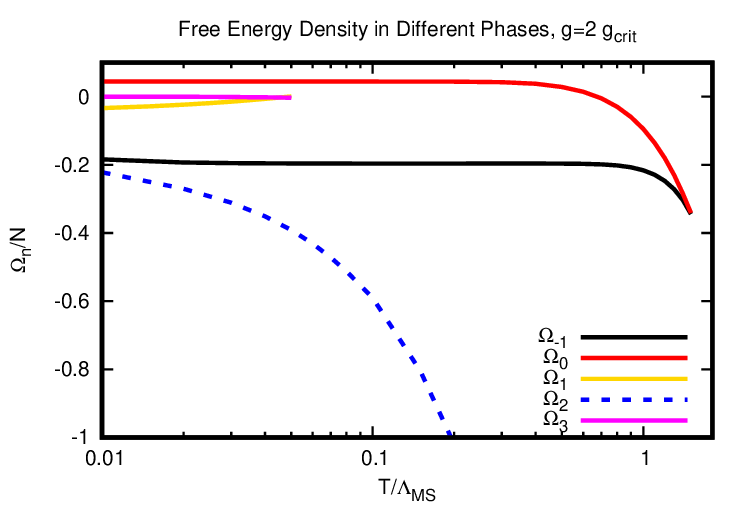}
  \caption{\label{fig:one} Free Energy Density for different phases of the negative coupling $O(N)$ model at large N. Shown are results for two different coupling values, $g=g_{\rm crit}$ and $g=2 g_{\rm crit}$, see text for details. Only cases with real-valued $\Omega_n$ are shown, so that absent or terminating lines imply that for these parameter values, the free energy is complex. The exception is $\Omega_2$ which is always found to be real-valued, for all parameter values.}
\end{figure}

Solving (\ref{saddle4}) numerically, one finds that it is the solution $m_{2}$ that is analytically connected to the high-temperature saddle (\ref{Rzhigh}). Unlike the solution connected to $i\bar z_{-1}$ from (\ref{zeroTsolutoins}), $m_{2}$ is always real for all temperature values. Defining $\Omega_n\equiv \Omega[m=m_n]$ as the free-energy density evaluated on the various saddles, Fig.~\ref{fig:one} shows the behavior of the free energies as a function of temperature for two cases of the coupling, $g=g_{\rm crit}$ and $g=2 g_{\rm crit}$. As can be seen from this figure, the saddle point $m=m_2$ typically provides the lowest free energy solution, except for the case of $g=g_{\rm crit}$ for temperatures below $T\lesssim 0.03 \Lambda_{\overline{\rm MS}}$, where $m=m_3$ is found to be the lowest free energy phase.

\subsection{NLO corrections at low temperature}

It is interesting to study the next-to-leading-order (NLO) corrections in the the large N expansion for these saddle points. The NLO correction to the free energy may be calculated by expanding (\ref{action}) to second order in fluctuations $\xi(x)$, cf. Ref.~\cite{Romatschke:2023ztk} for a pedagogical explanation. One finds
\be
S_{\rm eff}=S_{\rm eff}[z_0]+\frac{1}{2}\int d^2x \xi(x) D^{-1}(x-y) \xi(y)\,,
\ee
with the inverse auxiliary field propagator given by
\be
D^{-1}(x)=-\frac{N}{8g}+N \Pi(x)\,,\quad \Pi(x)\equiv \frac{G^2(x)}{2}\,,\quad G(\omega_n,k)=\frac{1}{\omega_n^2+k^2+m_B^2+i z_0}
\ee
and $G(x)$ the propagator for a single field $\phi(x)$.

In the zero temperature limit, after shifting $z_0$ as above, one finds
\be
T\rightarrow 0: \quad \Pi(k)=\frac{{\rm arctanh}\left(\sqrt{\frac{k^2}{k^2+4 iz_0}}\right)}{2\pi \sqrt{k^2\left(k^2+4 iz_0\right)}}\,,
\ee
and the free energy to NLO in large N becomes
\be
\label{NLOO}
\frac{\Omega}{N}=\frac{(i z_0)^2-2 iz_0 m_B^2+m_B^2}{16 g}+\frac{i z_0}{8\pi}\left[\frac{1}{\varepsilon}+\ln\frac{\bar\mu^2e^{1}}{i\zeta_0}\right]+\frac{1}{2N}\int \frac{d^dk}{(2\pi)^d} \ln D^{-1}(k)\,,
\ee
with $d=2-2\varepsilon$ in dimensional regularization. For all Euclidean momenta one has
\be
\label{NLOin}
\Pi(k)<\Pi(0)=\frac{1}{8 \pi i z_0}\,,
\ee
so that
\be
D^{-1}(k)<\frac{1}{8g}\left(-1+\frac{g}{\pi m_n^2}\right)\,,
\ee
where the saddle-point solution to leading order in large N, Eqns.(\ref{lolli}),(\ref{lolli2}), have been used. One immediately verifies that there are no poles of the Euclidean propagator $D(k)$ as long as $\frac{\pi m_n^2}{g}<1$, which is the case for $m_{-1},m_1$ and $g>g_{\rm crit}$, but not for $m_0,m_2$. Therefore, the saddles $m_0,m_2$, which were found to be thermodynamically disfavored from the analysis of the free energy (see Fig.~\ref{fig:one}), are found to have tachyons, while the thermodynamically stable saddles do not.

The easiest way to evaluate (\ref{NLOO}) is to add and subtract $\frac{1}{2}\int_k D(k)\Pi(k)$ and rewrite
\be
\int \frac{d^dk}{(2\pi)^d} D(k)\Pi(k)=\frac{1}{2}\int \frac{d^dk}{(2\pi)^d} G(k)\Sigma(k)\,,\quad \Sigma(x)=D(x)G(x)\,,
\ee
see e.g. \cite[Eq.(A8)]{Romatschke:2019rjk}, so that 
\ba
\frac{\Omega}{N}&=&\frac{(i z_0)^2+m_B^4}{16 g}+\frac{i z_0}{8\pi}\left[-\frac{\pi m_B^2}{g}+\frac{1}{\varepsilon}+\ln\frac{\bar\mu^2e^{1}}{i\zeta_0}\right]+\frac{1}{2N}\int \frac{d^dk}{(2\pi)^d} \left[\ln D^{-1}(k)-N D(k)\Pi(k)\right]\nonumber\\
&&+\frac{1}{4}\int \frac{d^dk}{(2\pi)^d} G(k)\Sigma(k)\,.
\ea
Now using the result \cite[Eq.(74)]{Romatschke:2024yhx} that $\Delta \Sigma(k)=\Sigma(k)-\Sigma(0)$ is divergence free, I use the leading-order large N saddle point condition (\ref{saddle}) to rewrite
\be
\frac{1}{4}\int \frac{d^dk}{(2\pi)^d} G(k)\Sigma(k)=-\frac{iz_0}{8g}\Sigma(0)+\frac{1}{4}\int \frac{d^dk}{(2\pi)^d} G(k)\Delta\Sigma(k)\,,
\ee
which leads to
\ba
\frac{\Omega}{N}&=&\frac{(i z_0)^2+m_B^4}{16 g}+\frac{i z_0}{8\pi}\left[-\frac{\pi m_B^2}{g}-\frac{\pi \Sigma(0)}{ g}+\frac{1}{\varepsilon}+\ln\frac{\bar\mu^2e^{1}}{i\zeta_0}\right]\nonumber\\
&&+\frac{1}{2}\int \frac{d^2k}{(2\pi)^2} \left[\ln D^{-1}(k)-N D(k)\Pi(k)\right]+\frac{1}{4}\int \frac{d^2k}{(2\pi)^2} G(k)\Delta\Sigma(k)\,,
\ea
where the two integrals in the last line are both free of UV-divergencies and can thus be evaluated numerically. To evaluate $\Sigma(0)$, I use appendix A in Ref.~\cite{Romatschke:2024yhx} to find
\be
T=0,\ g=2g_{\rm crit}:\quad \Sigma(0)=-\frac{2 g}{N \pi}\left(\frac{1}{\varepsilon}-\ln \frac{m_n^2e^{\gamma_E}}{\bar\mu^2}+0.019\right)\,,
\ee
where $m_n$ is the value of the saddle-point solution to leading order at large N, cf. Eqns.~(\ref{lolli}),(\ref{lolli2}). Therefore, the renormalization condition (\ref{renorm}) to NLO in large N must be
\be
\label{renormNLO}
\frac{m_B^2}{8g}-\frac{1+\frac{2}{N}}{8\pi \varepsilon}=\frac{m_R^2(\bar\mu)}{8g}\,,
\ee
as expected.

\subsection{NLO corrections at high temperature}

At high temperature, $\Pi(x)$ is dominated by the zero-frequency Matsubara mode. After shifting $z_0$ as before one has
\be
T\rightarrow \infty: \quad \Pi(\omega_n,k)=\frac{T}{2}\frac{1}{m_n\left(k^2+4 m_n^2\right)}\,,
\ee
where again $m_n$ denotes the value of the saddle-point solution at cf. Eqns.~(\ref{lolli}),(\ref{lolli2}). Using a similar inequality as for zero temperature (\ref{NLOin}), one has
\be
D^{-1}(k)<\frac{1}{8g}\left(-1+\frac{T g}{m_n^3}\right)\,.
\ee
As a consequence, this relation implies that there are poles of $D(k)$, and hence no tachyons, for the saddle-point solution $m_2=-(2 T g)^{\frac{1}{3}}$ at high temperature. 

\section{Conclusions}

In this work, I considered the two-dimensional O(N) model with upside-down quartic interactions. In the large N limit, I showed that at zero temperature the model can be non-perturbatively renormalized and is characterized by its saddle points which can be recognized as the pole-mass squared of the scalar propagator. For sufficiently strong negative coupling, the saddle points are real, and the largest pole mass corresponds to a physically sensible vacuum configuration.

At sufficiently high temperature, these saddle points cease to be real-valued, mirroring the situation encountered for the O(N) model in four dimensions \cite{Romatschke:2022jqg}. At very high temperature, the length of the thermal circle shrinks to zero, and the physics is capture by an effective theory with negative coupling which is only one-dimensional (quantum mechanics). I went on to demonstrate that the negative-coupling quantum mechanics theory can be recast into an equivalent Hermitian (positive-coupling) theory, which is easy to solve in the large N limit, but can also be solved numerically for any value of components N. By comparing the results from this quantum mechanical calculation with the saddle-point solution of the two-dimensional field theory in the high temperature limit, I showed that the results match if saddle-points on the non-principal Riemann sheet are allowed.

Armed with this insight, I recast the effective action of the large N two-dimensional quantum field theory in a form that is suitable to analytic continuation beyond the principal Riemann sheet, and found saddle points in addition to the ones on the principal Riemann sheet. Evaluating the free energy for all identified saddle points, I showed that the lowest free energy solutions are generically given by a saddle point on a non-principal Riemann sheet. Finally, evaluating corrections of the large N effective action to next-to-leading order in large N, I showed that the lowest free energy solution also is dynamically stable, whereas other saddle point solutions possess tachyons.

The possibility of saddle points contributing from non-principal Riemann sheets has been considered before \cite{Lawrence:2024pjg}, but the present calculation provides an example where this is actually \textit{necessary} in order to capture the dominant physics of the theory.

\section{Acknowledgments}

This work was supported by the Department of Energy, DOE award No DE-SC0017905. PR would like to thank Scott Lawrence for helpful discussions.

\bibliography{bib}
\end{document}